\documentclass[twocolumn,showpacs,preprintnumbers,amsmath,amssymb,floatfix,elsart]{revtex4}

\usepackage{graphicx}
\usepackage{dcolumn}
\usepackage{bm}

\newcommand{\beq}{\begin{equation}}
\newcommand{\eeq}{\end{equation}}
\newcommand{\bey}{\begin{eqnarray}}
\newcommand{\eey}{\end{eqnarray}}

\begin{document}

\title{Noncommutative geometry inspired wormholes with conformal motion}

\author{Farook Rahaman}
\email{ rahaman@iucaa.ernet.in} \affiliation{Department of
Mathematics, Jadavpur University, Kolkata 700032, West Bengal,
India}

\author{Saibal Ray}
\email{saibal@iucaa.ernet.in} \affiliation{Department of Physics,
Government College of Engineering \& Ceramic Technology, Kolkata
700010, West Bengal, India}

\author{G.S. Khadekar}
\email{gkhadekar@yahoo.com} \affiliation{Department of
Mathematics, R.T.M. Nagpur University, Nagpur 440033, Maharastra,
India }

\author{ P.K.F. Kuhfittig}
\email{kuhfitti@msoel.edu} \affiliation{ Department of
Mathematics, Milwaukee School of Engineering, Milwaukee, Wisconsin
53202-3109, USA}

 \author{Indrani Karar}
\email{indrani.karar08@gmail.com} \affiliation{Department of
Mathematics, Saroj Mohan Institute of Technology, Guptipara, West
Bengal, India.}

\date{\today}

\begin{abstract}
This paper discusses a new wormhole solution that admits
conformal motion, given a noncommutative-geometry background.
After a discussion of the wormhole geometry and the energy
conditions, the analysis proceeds with the calculation of the
active gravitational mass, a discussion of the TOV equation
describing the equilibrium conditions, as well as the nature
of the total gravitational energy.  The wormhole spacetime
is  not asymptotically flat and is therefore cut off at some
radial distance and joined to an exterior vacuum solution.
\end{abstract}


\maketitle

\section{Introduction}
Noncommutative geometry aims to unify General Relativity and the
Standard Model on the same geometrical footing so that one can
describe gravity and the electro-weak and strong forces as
gravitational forces of a unified spacetime \cite{stephan2013}.
Therefore, in a more concise and direct presentation
noncommutativity refers to the discretization of spacetime due to
the commutator $[ x^\mu, x^\nu ] = i \theta^{\mu \nu}$, where
$\theta^{\mu \nu}$ is an antisymmetric matrix. It is based on the
realization that in string theory coordinates may become
noncommutative operators on a $D$-brane \cite{eW96,SW99}.
Noncommutativity replaces point-like structures by smeared objects
\cite{NSS06, NS10, LG04, eS09} and is intended to eliminate the
divergences that normally appear in General Relativity. It is an
intrinsic property of spacetime and does not depend on any
particular features such as curvature.

A standard way to model the smearing effect is by using a Gaussian
distribution of minimal length $\sqrt{\theta}$ due to the
uncertainty. So the energy density of the static, spherically
symmetric and particle-like gravitational source has the form
\cite{NSS06, mR11}

\begin{equation}\label{source}
\rho =
\frac{M}{(4\pi\theta)^{\frac{3}{2}}}e^{-\frac{r^2}{4\theta}},
\end{equation}
where $M$ is the total mass of the source which here could be
considered as a diffused centralized object such as a wormhole
\cite{LG04}.

A number of studies inspired by noncommutative geometry can be
found in the literature \cite{RKCUR12,GL09,pK12}. Thus Rahaman et
al. \cite{RKCUR12} showed that a noncommutative-geometry
background is able to account for producing stable circular
orbits, as well as attractive gravity, without any need for exotic
dark matter. Garattini and Lobo \cite{GL09} obtained a
self-sustained wormhole in noncommutative geometry. Under their
semi-classical approach the energy density of the graviton one
loop contribution to a classical energy in a traversable wormhole
background and the finite one loop energy density is considered as
a self-consistent source for these wormhole geometries. Kuhfittig
\cite{pK12} showed that a special class of thin-shell wormholes
possess small regions of stability around a thin shell, even
though they are unstable in classical general relativity. Some
other works on wormholes, in connection with their origin and
development (with or without noncommutative-geometry), can be
looked at in the following references \cite{U10,RRI13}. On the
other hand, for Lorentzian traversable wormhole, where it would
allow travel in both directions from one part of the spacetime of
this universe or the another universe, one has to go through the
seminal papers of Thorne and his coworkers \cite{MT1988,MTY1988}.

Our starting point is the search for a natural relationship
between the wormhole geometry and the matter source supporting
such a wormhole, with the help of the Einstein field equations. So
we turn to the well-known inheritance symmetry contained in the
set of conformal Killing vectors (CKV) \cite{MM90, HP85, MMT86,
Ray2008, eS01}

\begin{equation}
        L_\xi g_{ik} =\xi_{i;k}+ \xi_{k;i} = \psi g_{ik}.
         \label{CKV}
          \end{equation}
\noindent Here $L$ is the Lie derivative operator and $\psi$ the
conformal factor. The vector $\xi$ generates the conformal
symmetry, while the metric tensor $g_{ik}$ is conformally mapped
into itself along $\psi$. Over time, CKV's have provided ever
deeper insights into the spacetime geometry. In the process these
conformal symmetries assist in generating exact solutions of the
Einstein field equations that may, in turn, be used to model
various relativistic, astrophysical, and cosmological phenomena
\cite{MS94, MM96,MH04, fR08}.

To obtain the desired connection to the relativistic as well as
astrophysical modeling of a wormhole in the framework of
noncommutative geometry, our principal aim is the construction of
a stable configuration fueled by a suitable amount of exotic
matter. To this end, we have introduced conformal Killing vectors
and basic equations in Sec. II. The important properties of the
wormhole thus obtained have been examined in Sec. III, including
studies on (i) active gravitational mass, (ii) TOV equation, and
(iii) total gravitational energy, explored and discussed in
several Sub-sections. A succinct summary of these results is
provided in Sec. IV.

\section{Conformal killing vectors and basic equations}
In this section we turn our attention to traversable wormholes,
starting with the static spherically symmetric metric

\begin{equation}
               ds^2=- e^{\nu(r)} dt^2+e^{\lambda(r)} dr^2
               +r^2( d\theta^2+\text{sin}^2\theta\, d\phi^2).
         \label{Eq3}
          \end{equation}

\noindent Here $\nu$ and $\lambda$ are the metric potentials of
the spacetime and function of radial coordinate $r$ only.

The Einstein field equations are stated next:

\begin{equation}\label{Einstein1}
e^{-\lambda}
\left[\frac{\lambda^\prime}{r} - \frac{1}{r^2}
\right]+\frac{1}{r^2}= 8\pi \rho,
\end{equation}

\begin{equation}\label{Einstein2}
e^{-\lambda}
\left[\frac{1}{r^2}+\frac{\nu^\prime}{r}\right]-\frac{1}{r^2}=
8\pi p_r,
\end{equation}

\noindent and

\begin{equation}\label{Einstein3}
\frac{1}{2} e^{-\lambda} \left[\frac{1}{2}(\nu^\prime)^2+
\nu^{\prime\prime} -\frac{1}{2}\lambda^\prime\nu^\prime +
\frac{1}{r}({\nu^\prime- \lambda^\prime})\right] =8\pi p_t,
\end{equation}

\noindent where $\rho$, $p_r$ and $p_t$ are respectively the
matter-energy density, radial and tangential pressures of the
fluid distribution.

Eq. (\ref{CKV}) now implies the following:

$$ \xi^1 \nu^\prime =\psi; ~~~ \xi^4  = C_1 = ~\text{constant};$$

\begin{equation}
\xi^1  = \frac{\psi r}{2}~~~ \text{and} ~~~ \xi^1 \lambda ^\prime
+ 2 \xi^1 _{,1}   =\psi.
                \label{Eq3}
\end{equation}

These, in turn, imply that

\begin{equation}
               e^\nu  =C_2^2 r^2,
     \label{gtt}
          \end{equation}

\begin{equation}
               e^\lambda  = \left(\frac {a} {\psi}\right)^2,
     \label{grr}
          \end{equation}

          and

\begin{equation}
               \xi^i = C_1 \delta_4^i
               + \left(\frac{\psi r}{2}\right)\delta_1^i,
       \label{Eq3}
          \end{equation}

where $C_2$ and $a$ are integration constants. We note from the
Eqs. (\ref{gtt}) and (\ref{grr}) that to get the explicit forms of
the metric potentials we have to fix the values of the constants
$C_2$ and $a$, and also the conformal factor $\psi$. However, to
find out solvable pattern of the Einstein equations
(\ref{Einstein1})-(\ref{Einstein3}) we can, at first hand,
transform these in terms of $\psi$. Thus using solutions
(\ref{gtt}) and (\ref{grr}), Eqs.
(\ref{Einstein1})-(\ref{Einstein3}) take the forms as follows:

\begin{equation}\label{alternate1}
\frac{1}{r^2}\left[1 - \frac{\psi^2}{a^2}
\right]-\frac{2\psi\psi^\prime}{ra^2}= 8\pi \rho,
\end{equation}

\begin{equation}\label{alternate2}
\frac{1}{r^2}\left[1 - \frac{3\psi^2}{a^2}
\right]= - 8\pi p_r,
\end{equation}

\begin{equation}\label{alternate3}
\frac{\psi^2}{a^2r^2}
+\frac{2\psi\psi^\prime}{ra^2} =8\pi p_t.
\end{equation}

At this point, it is now necessary to assume relationships between
the physical parameters $\rho$, $p_r$ and $p_t$. There are two
possibilities: either (1) to chose a suitable form of $\rho$ or
(2) to chose a suitable relation between $p_r$ and $p_t$. We
would, however, following Nicolini et al. \cite{NSS06} like to
have the first choice as will be shown below.

\section{Model of a wormhole and its properties}
It has been argued by several authors
\cite{Smailagic2003a,Smailagic2003b} that noncommutative geometry
offers conjecture of smeared or diffused objects to eliminate
point-like structures. To get this smearing effect we choose the
mass density of a static, spherically symmetric, diffused,
particle-like gravitational source as provided in Eq.
(\ref{source}).

Thus, using Eq. (\ref{source}), Eq. (\ref{alternate1}) becomes

\begin{equation}
\frac{1}{r^2}\left[1 - \frac{\psi^2}{a^2}
\right]-\frac{2\psi\psi^\prime}{ra^2}=  8 \pi \left(
\frac{M}{(4\pi\theta)^{\frac{3}{2}}}e^{-\frac{r^2}{4\theta}}
\right).
\end{equation}

Solving this equation, we get

\begin{equation}
\psi^2 = a^2-\left (\frac{4 M a^2 }{r \sqrt{4 \pi \theta}}\right
)\left[ -r   e^{-\frac{r^2}{4\theta}} + \sqrt{\theta \pi}~
erf\left(\frac{r}{2 \sqrt{\theta}}\right)\right] + \frac{D}{r},
\end{equation}

\noindent where $D$ is an integration constant.  Since $\psi$
exists at $r=0$, we let $D=0$.  As a result, we get the exact
analytical form for all the parameters, stated next:

\begin{equation}
 p_t  = \frac{1}{8 \pi} \left[ \frac{1}{r^2}
 - \frac{8 \pi M }{(4 \pi \theta)^{\frac{3}{2}}}
  e^{-\frac{r^2}{4\theta}}\right],
\end{equation}

\begin{equation}
 p_r  =-\frac{1}{8 \pi r^2} \left[1- 3\left\{ 1-\left
 (\frac{4 M  }{r \sqrt{4 \pi \theta}}\right
)\left( -r   e^{-\frac{r^2}{4\theta}} + \sqrt{\theta \pi}~
erf\left(\frac{r}{2 \sqrt{\theta}}\right)\right)\right\} \right],
\end{equation}

\begin{equation}
               e^\nu  =C_2^2 r^2,\label{Eq3}
\end{equation}

and

\begin{equation}
e^\lambda  = \frac{1}{1-\left (\frac{4 M }
               {r \sqrt{4 \pi \theta}}\right
)\left(-r   e^{-\frac{r^2}{4\theta}} + \sqrt{\theta \pi}~
erf\left(\frac{r}{2 \sqrt{\theta}}\right)\right)}. \label{Eq3}
\end{equation}

If the metric coefficient $e^{\lambda}$ is written in
terms of shape function $b(r)$, i.e., if

\begin{equation}
e^\lambda = \frac{1}{1-\frac{b(r)}{r}}, \label{Eq3}
\end{equation}

then $b(r)$ takes the form

\begin{equation}
b(r)= \left (\frac{4 M}{\sqrt{4 \pi \theta}}\right )\left[-r
e^{-\frac{r^2}{4\theta}} + \sqrt{\theta \pi}~ erf\left(\frac{r}{2
\sqrt{\theta}}\right)\right]. \label{Eq3}
\end{equation}

We have assumed the particle-like gravitational source in Eq.
(\ref{source}), which is positive (Fig. 1) and therefore results
in a shape function that is monotone increasing, so $b'(r)>0$
(Fig. 2). The throat of the wormhole is located at $r=r_0$, where
$b(r)-r$ cuts the $r$-axis, shown in Fig. 3. We also observe that
for $r>r_0$, $b(r)-r<0$, which implies that $b(r)/r<1$, an
essential requirement for a shape function.  Finally, since
$b(r)-r$ is decreasing for $r>r_0$, $b'(r_0)<1$, which is the
flare-out condition.

\begin{figure}
        \includegraphics[scale=.30]{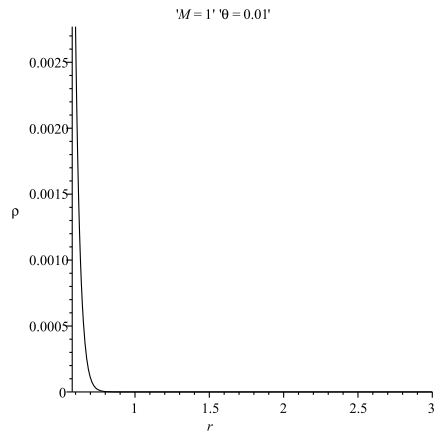}
        \caption{Plot of the energy density as a
        function of $r$.}
   \label{fig:wh1}
\end{figure}

\begin{figure}
        \includegraphics[scale=.30]{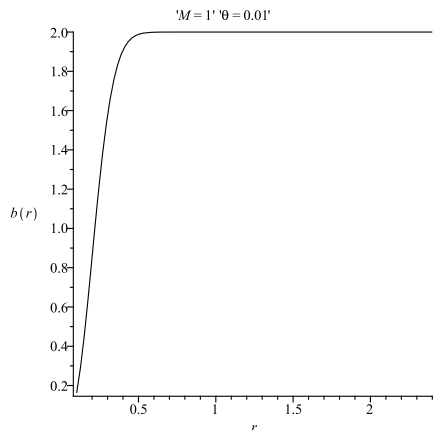}
        \caption{Diagram of the shape function of the
        wormhole  for specific values of the parameters:
        $\theta =0.01$, $M=1$. }
   \label{fig:shape1}
\end{figure}

One can note that Fig. 4 shows the asymptotic behavior
$b(r)/r\rightarrow 0$ as $r\rightarrow \infty$. Unfortunately, the
redshift function does not approach zero as $r\rightarrow\infty$
due to the conformal symmetry. So the wormhole spacetime is not
asymptotically flat and will therefore have to be cut off at some
radial distance and joined to an exterior vacuum solution.

\begin{figure}
        \includegraphics[scale=.30]{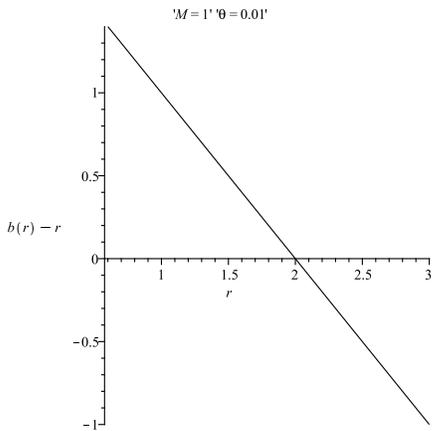}
        \caption{The throat of the wormhole, given in Fig. 2,
        occurs where $b(r)-r$ cuts the $r$-axis.}
   \label{fig:wh2}
\end{figure}

\begin{figure}
        \includegraphics[scale=.30]{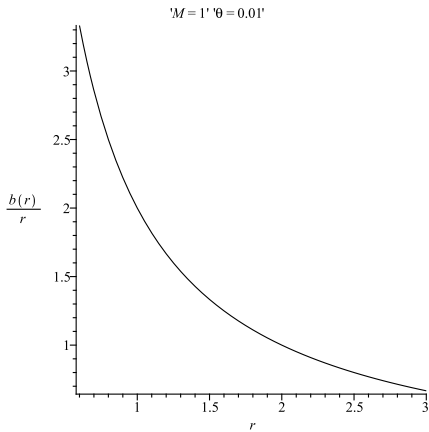}
        \caption{Asymptotic behavior  of the shape function
        of the wormhole given in Fig. 2.}
   \label{fig:shape2}
\end{figure}

From Eq. (\ref{gtt}), the redshift function is given by

\begin{equation}
   \frac{1}{2}\nu(r)=\text{ln}(C_2r).
\end{equation}

So we match our interior solution to the exterior Schwarzschild
solution at a junction interface $r=R$. Using this matching
condition, we can determine the constant $C_2$:

\begin{equation}
    C_2=\frac{e^{\nu(R)/2}}{R}.
\end{equation}

So, obviously the redshift function is finite in the region $r_0<
r <R$, as required.

Finally, let us check whether the material threading the wormhole
violates the energy conditions. According to Fig. 5, the Null
Energy Condition (NEC) and the Weak Energy Condition (WEC) are
violated, conditions that are necessary to hold a wormhole open.
It is interesting to note that the Strong Energy Condition (SEC)
is met away from the throat. By contrast, when dealing with
noncommutative wormholes that do not admit conformal motion, the
SEC is violated \cite {eS09}.

\begin{figure}
        \includegraphics[scale=.30]{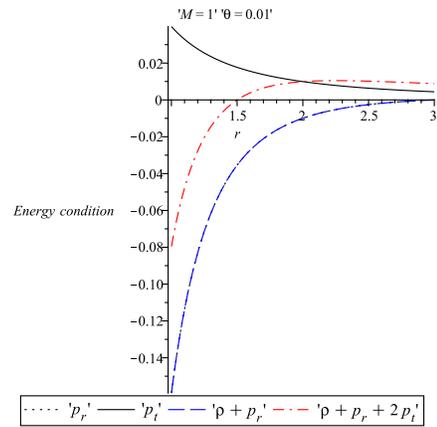}
        \caption{Variation of the left-hand sides of the
        respective expressions for the energy conditions
        plotted against $r$.}
   \label{fig:wh20}
\end{figure}

\subsection{Active gravitational mass}
Our next step is to calculate the active gravitational mass within
the region from $r_0$ up to the radius $R$ \cite{Varela2010,
Rahaman2010}:

\begin{equation}
\label{eq40} M_{active}=4\pi\int_{r_0}^{a} \rho r^2 dr\\
= M  \left [\frac{-2r  e^{-\frac{r^2}{4\theta}}+2\sqrt{\pi}
\,\theta^{\frac{3}{2}} erf \left( \frac{r}{2 \sqrt{\theta}}
\right)}{2  \sqrt{\pi }\, \theta^{\frac{3}{2}}} \right]^a_{r_0}.
\end{equation}

Observe that $M_{active} \longrightarrow M$ (Fig. 6). So a distant
observer would not be able to distinguish the gravitational nature
of the wormhole from a compact mass $M$. What needs to be
emphasized is that Eq. (\ref{eq40}) gives the correction due to
the noncommutative geometry.

\begin{figure}
        \includegraphics[scale=.30]{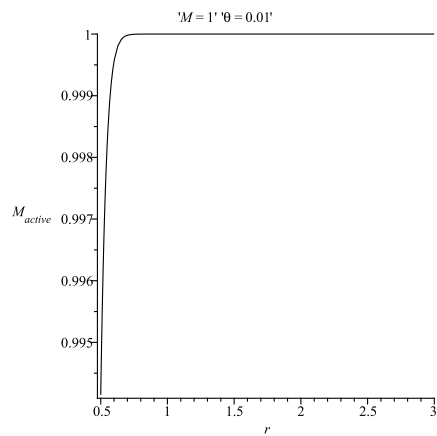}
        \caption{The active gravitational mass
$M_{active}$ plotted against $r$.}
   \label{fig:wh20}
\end{figure}

\subsection{The Tolman-Oppenheimer-Volkoff (TOV) equation}
At this point let us consider the equilibrium stage that can be
achieved for matter threading the wormhole. To that end, we write
the TOV equation as derived by solving the Einstein equations for
a general time-invariant, spherically symmetric metric in the
following form \cite {Varela2010, Rahaman2010, Leon1993}:

\begin{equation}
-\frac{M_G\left(\rho+p_r\right)}{r^2}e^{\frac{\lambda-\nu}{2}}
 -\frac{dp_r}{dr}
 +\frac{2}{r}\left(p_t-p_r\right)=0, \label{TOV}
\end{equation}

\noindent where $M_G=M_G(r)$ is the effective gravitational mass
within the region (from $r_0$ up to the radius $r$), and is given
by

\begin{equation}
M_G(r)=\frac{1}{2}r^2e^{\frac{\nu-\lambda}{2}}\nu^{\prime}.
   \label{eq32}
\end{equation}

The TOV equation (\ref{TOV}) describes the equilibrium condition
for the wormhole, subject to the gravitational force $F_g$ (due to
gravitating mass) , the hydrostatic force $F_h$ (due to
hydrostatic fluid), and the anisotropic force $F_a$ (due to
anisotropy of the system). To study the equilibrium of the
physical system, we write Eq. (\ref{TOV}) suitably in the
following form

\begin{equation}
 F_g+ F_h + F_a=0,\label{eq33}
\end{equation}

\noindent where

$$ F_g =-\frac{\nu^\prime}{2}\left(\rho +p_r\right)~~~~~~~~~~~~~~~~~~~~~~~~~~~~~~~~~~~~~~~~~~~~~~~~~~~~~~~~~~~~~~~~~~~~$$

$$= \frac{1}{8 \pi r^3} \left[1- 3\left\{ 1-\left (\frac{4 M  }{r
\sqrt{4 \pi \theta}}\right )\left( -r   e^{-\frac{r^2}{4\theta}} +
\sqrt{\theta \pi}~ erf\left(\frac{r}{2
\sqrt{\theta}}\right)\right)\right\} \right]$$

\begin{equation}
-\frac{M}{r(4\pi\theta)^{\frac{3}{2}}}e^{-\frac{r^2}{4\theta}},~~~~~~~~~~~~~~~~~~~~~~~~~~~~~~~~~~~~~~~~~~~~~~~~~~~~~~~~~
\label{eq34}
\end{equation}

$$ F_{h}= -\frac{dp_{r}}{dr}= \frac{1}{2 \pi r^{3}}
+\left(\frac{3M}{\sqrt{ 4\pi\theta}}\right)
\left(\frac{e^{-\frac{r^2}{4\theta}}}{r^{3}}\right)
+\left(\frac{3M}{2\pi\sqrt
{4\pi\theta}}\right)e^{-\frac{r^2}{4\theta}}$$

\begin{equation}
 +\left(\frac{3M}{4 \theta\sqrt {4\pi\theta}}\right)\left(\frac{e^{-\frac{r^2}{4\theta}}}{r}\right)
 -\left(\frac{9M\sqrt {\theta \pi}}{2\pi \sqrt {4\pi \theta}}\right)\left(\frac{erf (\frac{r}{2\sqrt\theta})}{r^4}\right),
\end{equation}

\noindent and

 $$ F_{a}= \frac{2}{r}(p_{t}-p_{r}) = \frac{1}{4\pi r} \left[  \frac{1}{r^{2}}-
 \frac{8\pi M }{(4\pi \theta)^{\frac{3}{2}}} e^{-\frac{r^2}{4\theta}} \right]$$

\begin{equation}
 +\frac{1}{4\pi r^{3}}\left[1- 3 \left\{1- \left(\frac{4M}{r\sqrt{4\pi \theta}}\right)
 \left(  -r e^{-\frac{r^2}{4\theta}}+ \sqrt{\theta \pi}\;\;erf (\frac{r}{2\sqrt\theta})\right)\right \}\right].
\end{equation}

The profiles of $F_g$, $F_h$, and $F_a$ are shown in Fig. 7. It
can be observed that, among the forces gravitational one is weak
relative to the others and counter balancing are mainly taking
part between hydrostatic and anisotropic forces to achieve
equilibrium stage out side the throat. Thus Fig. 7 distinctly
indicates that the overall equilibrium can be achieved due to the
combined effects of gravitational, hydrostatic, and anisotropic
forces, as a consequence of the conformal motion and the
noncommutative geometry.

\begin{figure}[htbp]
\centering
\includegraphics[scale=.30]{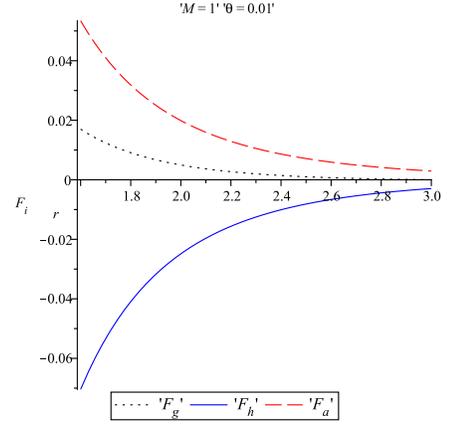}
\caption{Three different forces acting on the fluid elements
in static equilibrium plotted against $r$.\label{fig5}}
\end{figure}

\subsection{Total gravitational energy}
We have already observed that the material composing the wormhole
violates the NEC and WEC, and must therefore be of the exotic
rather than normal type. The total gravitational energy of a
structure composed of normal baryonic matter is negative. So it is
important to determine the nature of the gravitational energy in a
wormhole setting.

\begin{figure}[ptb]
\begin{center}
\vspace{0.3cm} \includegraphics[width=.30\textwidth]{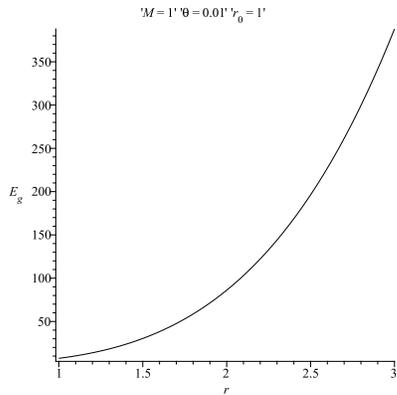}
\end{center}
\caption{Plot of $E_g$ as a function of $r$. } \label{fig4}
\end{figure}

Lynden-Bell et al. \cite{Lyndell} prescribed the formula for the
total gravitational energy of any stationary spacetime
(considering absence of black holes) as follows: $E_g = M - E_M$,
where $M$ is the total mass as defined earlier and $E_M$ is the
gravitational binding energy. However, following Nandi et al.
\cite{Nandi}, we can obtain an expression for this total
gravitational energy in it's explicit form, applied particularly
to a wormhole:

\begin{equation}
\label{eq41} E_g= M - E_M = \frac{1}{2}  \int_{r_0}^r [ ~1
-\sqrt{g_{rr}}~] \rho r^2 dr +  \frac{r_0}{2}.
\end{equation}

\noindent Here $g_{rr} = (1- \frac{b(r)}{r})^{-1}$ as defined in
terms of shape function $b(r)$, and the factor $\frac{r_0}{2}$ is
associated with the effective gravitational mass. It can be argued
that since $\sqrt{g_{rr}} > 1$, then $E_g < 0$ (attractive) if
$T^0_0 > 0$ and that $E_g > 0$ (repulsive) if $T^0_0 < 0$
\cite{Misner,Nandi}. We are now interested to evaluate $E_g$ and
hence take the range of the integral to be from the throat $r_0$
to the embedded radial space of the wormhole geometry. So the
total gravitational energy of the wormhole is given by

\begin{equation}
\label{Eg} E_g= \left[-\frac{.0129 r^2}{\theta^{\frac{3}{4}}} +
\frac{.0037 r^3}{\theta^{\frac{3}{2}}} +\frac{.0011
r^4}{\theta^{\frac{7}{4}}} +0(r^5)\right]^r_{r_0} + \frac{r_0}{2}.
\end{equation}

\noindent Due to complicated nature of the integrand, we cannot
obtain an exact analytical form and must therefore approximate the
total gravitational energy graphically. The plot for total
gravitational energy, obtained by retaining terms up to fourth
order in Eq. (\ref{Eg}), indicates that $E_g > 0$. Note that we
get $E_g>0$ here in spite of $T_0^0>0$. This is because the matter
distribution that supported the wormhole structure violates the
NEC and WEC. In other words, there is a repulsion around the
throat (See Fig. 8). This repulsive nature or positivity of $E_g$
is very much expected in a physically valid wormhole.

\section{Concluding Remarks}
In this paper we investigate traversable wormholes admitting
conformal motion that are also supported by a
noncommutative-geometric matter source. Some of the important
features of the present investigation can be formulated as
follows:

(1) Since, due to the conformal symmetry, the spacetime is not
asymptotically flat, it was necessary to cut off the material and
to match the interior wormhole solution to an exterior
Schwarzschild spacetime at a junction interface $r=R$.

(2) We also calculated the active gravitational mass in the region
extending from $r=r_0$ to $r=R$, and then observed that at a
certain stage the active mass, $M_{active}$, tends to the total
mass, $M$, of the system. This effect then indicates that a
distant observer would not be able to distinguish between the
gravitational nature of the wormhole and a compact mass of the
source.

(3) This is followed by a discussion of the TOV equation
describing the effect of the conformal motion on the equilibrium
stage for the wormhole in terms of the gravitational force $F_g$,
the hydrostatic force $F_h$, and the anisotropic force $F_a$.

(4) It is shown that the total gravitational energy is positive,
attributable to the repulsion around the throat.

(5) Finally, in the framework of wormhole construction the matter
violating the NEC and WEC conditions plays a significant role to
avoid the tunnel to collapse. This type of matter clearly a sharp
departure from the behavior of normal matter. Therefore, our
intention is justified to consider Gaussian distribution which is
the building block of the matter distributions that supply fuel to
construct wormhole.

\section*{Acknowledgement}
FR and SR gratefully acknowledge support from IUCAA, Pune, India
for providing Visiting Associateship under which a part of this
work was carried out. FR is also thankful to PURSE for providing
financial support.

\pagebreak

\end{document}